# Performance Limits of Monolayer Transition Metal Dichalcogenide Transistors

Leitao Liu, S. Bala Kumar, Yijian Ouyang, and Jing Guo[(a)]

*Department of Electrical and Computer Engineering, University of Florida, Gainesville, FL, 32608, USA.*

ABSTRACT

The performance limits of monolayer transition metal dichalcogenide transistors are examined with a ballistic MOSFET model. Using *ab-initio* theory, we calculate the band structures of two-dimensional (2D) transition metal dichalco-genide ($MX_2$). We find the lattice structures of monolayer $MX_2$ remain the same as the bulk $MX_2$. Within the ballistic regime, the performances of monolayer $MX_2$ transistors are better compared to the silicon transistors if thin high-κ gate insulator is used. This makes monolayer $MX_2$ promising 2D materials for future nanoelectronic device applications.

## I. INTRODUCTION

Graphene, a 2D material with great electrical, thermal and mechanical properties, has been widely studied in the past several years [1][2]. However, due to the absence of band gap, the graphene based field-effect-transistors (FETs) have high off-currents. Therefore different approaches are proposed to induce band gap in graphene, e.g. lateral confinement into nanoribbons. But these approaches often result in a significant reduction of carrier mobility [3][4], loss of coherence [5], and increased off-current [6] compared to graphene.

$MX_2$ (M = Mo, W; X = S, Se, Te) belongs to the family of layered transition metal dichalcogenide, whose crystal structure is built up of X-M-X monolayers interacted through van der Waals forces. Each monolayer consists of two X atom layers and a M atom layer sandwiched between the X atom layers. Previous research found that $MX_2$ has a band gap of 1.1eV ~ 2.0eV [7][8]. Therefore, monolayer of semiconduct-ing $MX_2$, a 2D material, may be suitable for CMOS-like logic device applications, and may be a promising replacement for silicon (Si). Scotch-tape based micromechanical cleavage technique can be used to extract a stable single layer $MX_2$ [9][10]. And liquid exfoliation approaches have also been proposed to produce 2D nanosheets [11][12]. It has been successfully shown that layered materials, such as $MoS_2$, $MoSe_2$, $MoTe_2$, $WS_2$, $TaSe_2$,

(a) Corresponding author: guoj@ufl.edu.

NbSe$_2$, NiTe$_2$, BN and Bi$_2$Te$_3$, can be efficiently exfoliated into individual layers [12]. The research on electronic devices based on these novel 2D materials has been greatly advanced due to these exfoliation techniques. A first successful realization of a FET based on monolayer MoS$_2$ has been reported with both high carrier mobility and high on/off current ratio [13].

Although monolayer MX$_2$ based transistors have been demonstrated experimentally, there is lack of theoretical investigations of their device performances. Thus, a comprehensive study and comparison of performance limits of these transistors from a theoretical model would be essential. In this paper, we first investigate the band structures of monolayer MX$_2$ using *ab-initio* theory, and calculate their effective masses through fitting the band structures. Then, we examine the ballistic performance limits of MOSFET with these materials as the channel materials, and make a comparison with Si thin film (2D-Si) FETs.

## II. APPROACH

### A. Electronic Band Structure Calculation

Fig. 1 shows the atomic structure of MX$_2$ (M = Mo, W; X = S, Se, Te). MX$_2$ is a layered material which is composed of vertically stacked X-M-X layers through van der Waals forces. Each single X-M-X sandwich layer consists of two hexagonal planes of X atoms and an intermediate hexagonal plane of M atoms interacting through ionic-covalent interactions with a triangular prismatic coordination.

We employ the Vienna *ab-initio* simulation package (VASP) [14][15] to perform density functional theory (DFT) calculations, within the formalism of the projector augmented wave method [16]. A double zeta polarized (DZP) basis set is used. The generalized gradient approximation (GGA) corrected functional by Perdew-Burke-Ernzerhof [17] is used for the exchange-correlation potential. The cutoff energy for the wave function expansion is set to 500eV, and a mesh of $18 \times 18 \times 1$ *k* points is used for 2D Brillouin-zone integrations.

We calculate the band structure of monolayer MoS$_2$ using three different methods. In the first method, we perform a regular GGA calculation, where we relax the structure first, and then calculate band structure. The calculated band gap $E_g$ = 1.68eV. In the second method, we perform structure relaxation and band structure calculation using HSE06 [18], and obtain $E_g$ = 2.27eV. Compared to the reported experimental value of the band gap 1.8eV [19], above GGA and HSE06 calculations in the regular methods do not accurately predict the band gap of monolayer $MoS_2$. Therefore, in the third method we construct the lattice structure of monolayer $MoS_2$ by using the measured bulk $MoS_2$, and perform band structure calculation without structure relaxation. The calculated band gap is 1.78eV, which is in closer agreement with the experimental value compared to the previous two calculations. This indicates that the structure of monolayer $MoS_2$ is likely to remain the same as the bulk $MoS_2$ [10][20]. Therefore, we adopt the third method for band structure calculations. However, note that, till present there is no direct experimental evidence which indicates that the structure of a single-layer $MoS_2$ is indeed the same as the bulk. Based on our calculations, the band structures of monolayer $MX_2$ are shown in Fig. 2. The band gap of monolayer $MX_2$ occurs at the high-symmetry K point, and is direct, unlike the bulk $MX_2$ [19][21]. We calculate the effective masses by a parabolic fitting of the band structure along corresponding crystal directions. The calculation results for $MX_2$ are listed in Table 1.

### B. Ballistic Performance Limits

Here we use the analytical ballistic MOSFET model [25]-[29] to investigate the performance limits of transistors with monolayer $MX_2$ semiconductors as the channel materials. Fig. 3 illustrates the structure of the device used in the simulation, and Fig. 4(a) illustrates the potential barrier and the source and drain Fermi energy levels. At zero terminal bias, the equilibrium electron density at the top of energy barrier is

$$N_0 = \int_{-\infty}^{+\infty} D(E) f(E - E_F) dE \qquad (1)$$

where $D(E)$ is the density of states at energy $E$, and $f(E-E_F)$ is the Fermi distribution with $E_F$ as the Fermi level. When gate and drain biases are applied, the energy barrier is modulated accordingly. The positive velocity states at the top of the barrier are filled by electrons from the source, and the negative states are filled by electrons from the drain. The electron density is given by

$$N = \frac{1}{2} \int_{-\infty}^{+\infty} D(E - U_{scf}) \times \left[ f(E - E_{FS}) + f(E - E_{FD}) \right] dE \quad (2)$$

where $E_{FS}(E_{FD})$ is the Fermi level in the source (drain), and $U_{scf}$ is the self-consistent surface potential which is calculated by coupling the charge density calculation to a capacitance model that describes transistor electrostatics as shown in Fig. 4(b):

$$U_{scf} = U_L + U_P \tag{3}$$

$$U_L = -q\left(\alpha_G V_G + \alpha_D V_D + \alpha_S V_S\right) \tag{4}$$

$$U_P = \frac{q^2}{C_G + C_D + C_S}(N - N_0) \tag{5}$$

where $\alpha_G = C_G / (C_G + C_D + C_S)$, $\alpha_D = C_D / (C_G + C_D + C_S)$ and $\alpha_S = C_S / (C_G + C_D + C_S)$. Once the convergence is achieved, the ballistic current $I_{DS}$ can be evaluated through the difference between the flux from the source and drain. And then the average electron velocity $v_{avg}$ at the top of the barrier is then evaluated as

$$v_{avg} = \frac{I_{DS}}{Q} = \frac{I_{DS}}{qN} \tag{6}$$

Detailed derivations of this model are described in [27]. For comparison, we also include the simulated results of 2D-Si transistors at the same operating bias. The thickness of silicon thin film is set to 5nm, and hence three sub-bands are considered for n-type 2D-Si transistors.

III. RESULTS AND DISCUSSIONS

As shown in Fig. 3, here we use a double-gate MOSFET with the high-κ $ZrO_2$ ($\varepsilon_r = 25$) dielectric insulator thickness $t_{ins} = 3$nm, which results in a gate capacitance of $C_G = 0.1476$ pF/μm². But for 2D-Si transistors, since quantum mechanical effect in the direction normal to the interface between the channel and the gate insulator is significant, the average inversion layer actually locates away from the interface, increasing the total insulator layer thickness equivalently. The additional thickness of the insulator can be simply calculated as follows [30]:

$$\Delta t_{ins} = \frac{\varepsilon_{ins}}{\varepsilon_{Si}} \times \Delta x_{av} \tag{7}$$

where $\varepsilon_{ins}$ is the relative permittivity of the gate dielectric, $\varepsilon_{Si}$ is the relative permittivity of Si, and the difference of the depth of the inversion layer to the surface with and without quantum mechanical effect $\Delta x_{av}$

= 10-12Å [30]. This value is valid for a wide range of channel doping and effective fields. In addition, a thin $SiO_2$ is formed at the surface of Si, even when high-κ gate insulator material is used. In contrast, there are no dangling bonds and therefore no formation native oxide on the surface of monolayer $MX_2$. This effect results in a further increase in the total gate insulator thickness in 2D-Si transistors. Therefore, the gate control on the channel charge density for monolayer $MX_2$ transistors is stronger than 2D-Si transistors at any given biases. Due to the effects discussed above, here we use an equivalent $ZrO_2$ thickness of 5.5nm for 2D-Si transistors.

Before implementing the analytical model described above, first we identify the parameters $α_G$, $α_S$ and $α_D$. These parameters represent the controllability of corresponding terminal on the modulation of the energy barrier. For an ideal MOSEFT, the gate completely controls the potential, while the influence of the source and drain are negligible, i.e. $α_G ≈ 1$ and $α_S, α_D ≈ 0$. In a more realistic case, these parameters can be obtained by fitting the experimental results. Here we set $α_G = 0.88$ and $α_D = 0.035$, respectively. To compare the performances of transistors with different channel materials, we adjust $E_{FS}$ to achieve a fixed off-current density of 0.003μA/μm. The $E_{FS}$ can be adjusted by changing the work function of the gate and the doping density in the source. Note that the Fermi level used in our calculations (intrinsic Fermi level) of $MoS_2$, $MoSe_2$, $MoTe_2$, and $WS_2$ are -0.3406 (-3.5071), -0.3412 (-3.1357), -0.3414 (-2.3308), and -0.3340 (-3.2378) eV, respectively.

The calculated results of an n-type MOSFET along the $k_x$ direction are presented in Fig. 5. Fig. 5(a) and 5(b) show the $I_{DS}$-$V_G$ and $I_{DS}$-$V_D$ characteristics, respectively. All the monolayer $MoX_2$, i.e. $MoS_2$, $MoSe_2$ and $MoTe_2$ transistors have similar values for the on-current, and these values are slightly higher than that of the 2D-Si transistors. On the other hand, within the ballistic regime the monolayer $WS_2$ transistors have the best performance. $WS_2$ transistors outperform 2D-Si transistors in terms of on-current by about 28.3%. Due to the atomic body thickness of the monolayer $MX_2$, transistors with these materials exhibit good gate control and result in high on-current. This makes monolayer $MX_2$ to be promising channel materials to replace Si for future FETs. Fig. 6 shows the average velocity of electron versus the gate voltage at $V_D = 0.6V$. The electron transport effective mass in monolayer $MX_2$ is larger than that in 2D-Si, and hence the electron velocity in monolayer $MX_2$ transistors is lower than that in 2D-Si transistors. And the performances of all the different monolayer $MoX_2$ transistors are similar due to the similarity in the electron effective masses.

The calculated results of a p-type MOSFET along the $k_x$ direction is presented in Fig. 7 and Fig. 8. Here we consider both the heavy and light hole of Si. The p-type MOSFET shows similar conclusions as the n-type MOSFETs. The on-current of monolayer $WS_2$ transistors is 1.5 times larger than that of 2D-Si transistors. And monolayer $MoS_2$ transistors outperform 2D-Si transistors in terms of on-current by about 27.3%.

Note that we also evaluate ballistic performances along the other crystal direction, namely the $k_y$ direction, and find that the results are very close to those along the $k_x$ direction with a difference of less than 3%. Therefore, we have only presented the results for the $k_x$ direction.

Finally, we study the effect of using a thicker gate oxide on the transistor performance. For a thicker gate oxide, the decrease of the gate capacitance due to quantum effects and silicon native oxide are less important. For a $SiO_2$ gate insulator thickness of 10nm, the 2D-Si transistors deliver a 20.6% larger on-current than monolayer $WS_2$ transistors, because the larger carrier velocity has a more dominant effect than slight degradation of the gate capacitance in 2D-Si MOSFETs as discussed before. Therefore, to exploit the performance advantage of monolayer $MX_2$ transistors in terms of ballistic on-current, a thin high-κ gate insulator is necessary.

For a more practical model, the influence of phonon scattering, impurity scattering and contact resistance should be considered to provide a more complete picture, which is well beyond the simple model used in this paper.

## IV. Summary

In this paper, we have first calculated the band structures of monolayer $MX_2$ (M = Mo, W; X = S, Se, Te), which are new two-dimensional semiconductors that can be effectively produced. By comparing the calculated results using different methods, we find a fixed structure with bulk lattice parameters give results closest to the experimental data. This is because the structure of monolayer $MX_2$ remains the same as the bulk $MX_2$. Using this method, we further compute the energy band gaps and effective masses along different crystal directions. And then, we adopt a ballistic MOSFET model to evaluate the performance limits of monolayer $MX_2$ transistors, and compare the results with those of 2D-Si transistors. We find that the ballistic performances of monolayer $MoX_2$ transistors are very similar to each other, while monolayer $WS_2$ transistors have a better performance. All of them outperform 2D-Si transistors in terms of the ballistic on-current when thin high-κ gate insulator is applied. Monolayer MX$_2$ transistors can exhibit better gate controllability because of the atomic body thickness, and thus they are promising 2D materials for future nanoelectronic device applications.

**Note**: During reviewing of this paper, we become aware of a paper published by A. Kuc *et. al.* [31], which shows that the combination of a different functional and a relaxed structure, results in a similar band gap for the monolayer MoS$_2$.

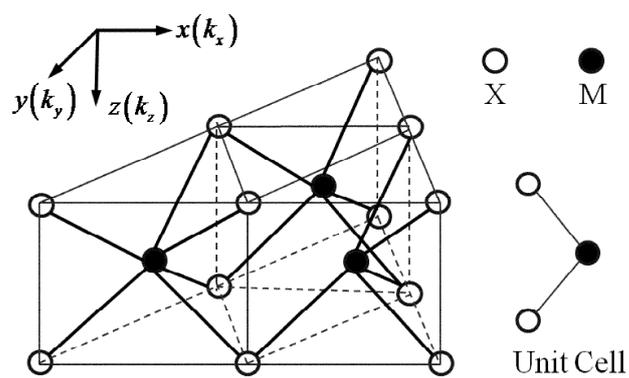

Fig. 1. Crystal structure of monolayer MX$_2$. The atoms form a trigonal prismatic coordination, and a hexagonal M atom layer is sandwiched between two hexagonal X atom layers. *a* is the lattice constant, and *c* is the height of X-M-X layer.

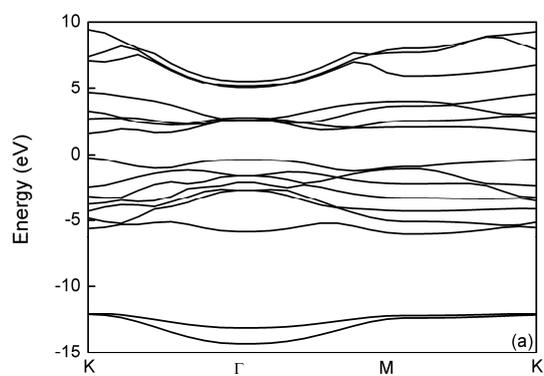

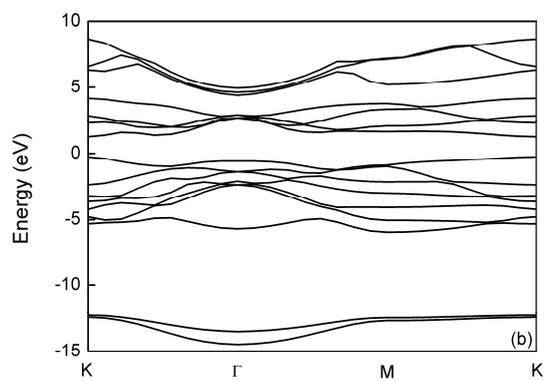

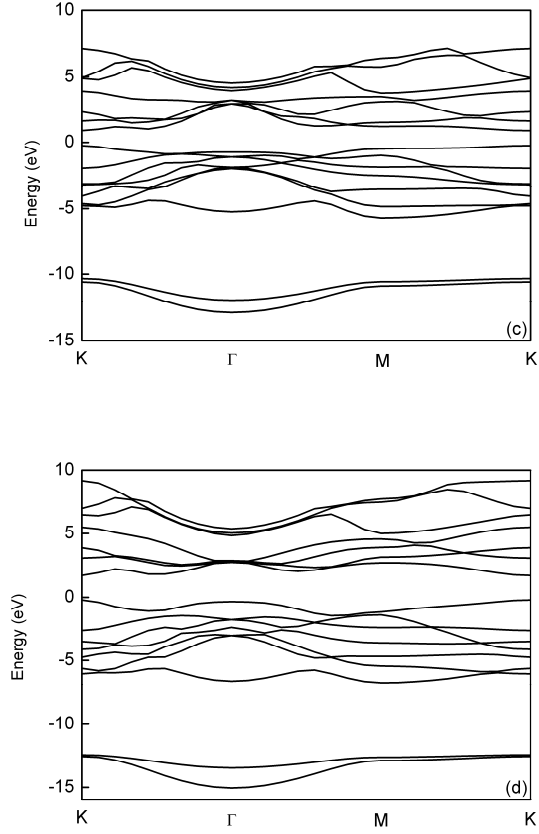

Fig. 2. The band structure of monolayer MX$_2$ calculated with a fixed crystal structure constructed by the bulk lattice parameters using GGA. The Fermi level is set at 0eV. These materials are direct semiconductors, and the band gap occurs at the high-symmetry K point. (a) monolayer MoS$_2$, $E_F$ = -3.5071eV. (b) monolayer MoSe$_2$, $E_F$ = -3.1357eV. (c) monolayer MoTe$_2$, $E_F$ = -2.3308eV. (d) monolayer WS$_2$, $E_F$ = -3.2378eV.

TABLE I
CALCULATED RESULTS OF MONOLAYER MX$_2$ USING GGA

|  | MoS$_2$ | MoSe$_2$ | MoTe$_2$ | WS$_2$ |
| --- | --- | --- | --- | --- |
| $a$ (Å) | 3.16[1] | 3.299[1] | 3.522[1] | 3.1532[1] |
|  | 3.182[2] | 3.311[2] | 3.539[2] | 3.181[2] |
| $c$ (Å) | 3.172[1] | 3.338[1] | 3.604[1] | 3.1424[1] |
|  | 3.138[2] | 3.348[2] | 3.622[2] | 3.140[2] |
| $E_g$ (eV) | 1.78[3] | 1.49[3] | 1.13[3] | 1.93[3] |
| $m_n^x$ (×$m_0$) | 0.5788[3] | 0.6059[3] | 0.6164[3] | 0.3466[3] |
| $m_n^y$ (×$m_0$) | 0.5664[3] | 0.5933[3] | 0.6033[3] | 0.3382[3] |
| $m_p^x$ (×$m_0$) | 0.6659[3] | 0.7114[3] | 0.7586[3] | 0.4619[3] |
| $m_p^y$ (×$m_0$) | 0.6524[3] | 0.6967[3] | 0.7406[3] | 0.4501[3] |

$E_g$ is the energy band gap. $m_n^x$ ($m_n^y$) and $m_p^x$ ($m_p^y$) is the fitting effective mass for electron and hole along the $k_x$ ($k_y$) direction, respectively.

[1] Experimental data of bulk lattice parameters [7][8][22][23][24]

[2] Calculated data of monolayer MX$_2$ lattice parameters by GGA with structure relaxation.

[3] Fitting parameters against band structure calculated by GGA with a fixed structure (using bulk lattice parameters in (1)).

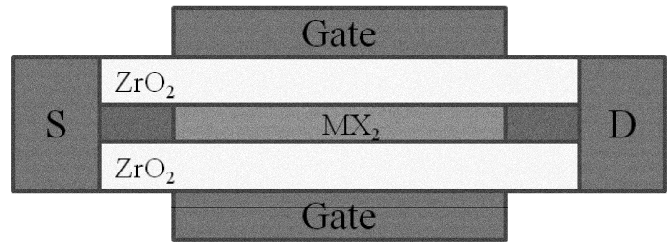

Fig. 3. Structure of a double-gate MOSFET model. The thickness of $ZrO_2$ dielectric insulator $t_{ins}$ = 3nm.

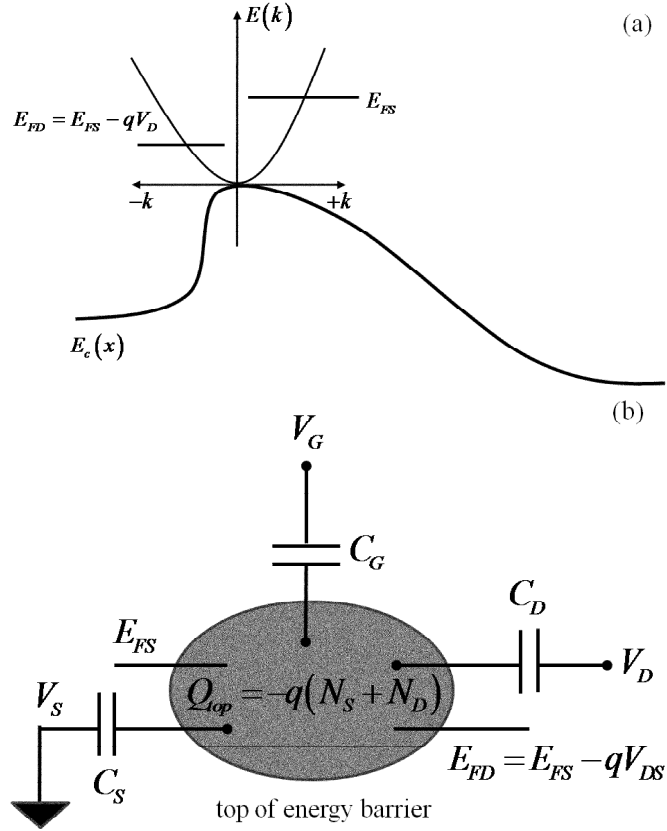

Fig. 4. (a) Illustration of the potential barrier and the source and drain Fermi energy levels. The $+k$ states are occupied by carrier from the source and the $-k$ states are occupied by carrier from the drain. (b) 2D transistor model of ballistic MOSFET. The potential at the top of the energy barrier is, $U_{scf}$, controlled by the gate, source and drain through the three capacitors $C_G$, $C_S$ and $C_D$, respectively.

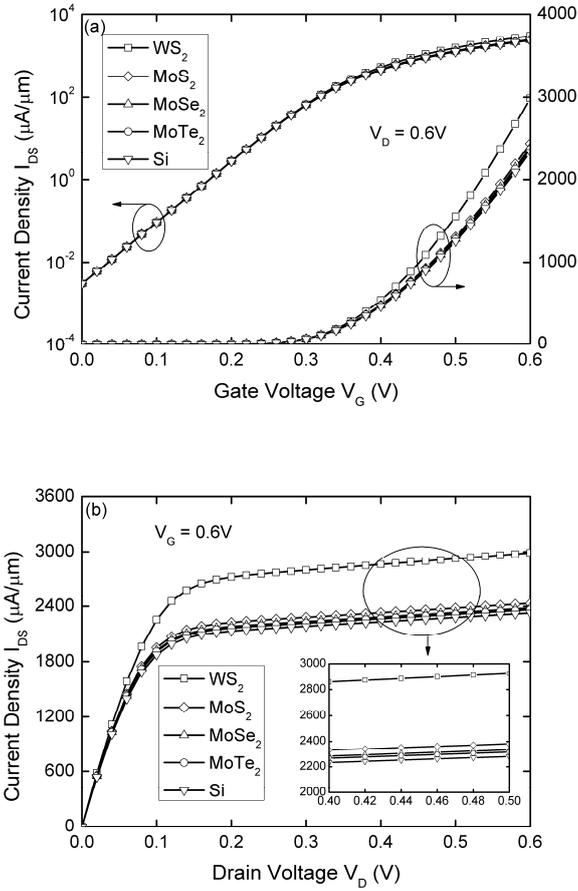

Fig. 5. Ballistic *I-V* characteristic comparison of n-type monolayer MX$_2$ transistors and 2D-Si transistors. Three sub-bands are considered in the calculation for Si thin film. The off-current is set to be 0.003μA/μm. (a) $I_{DS} \sim V_G$ with $V_D$ = 0.6V. (b) $I_{DS} \sim V_D$ with $V_G$ = 0.6V.

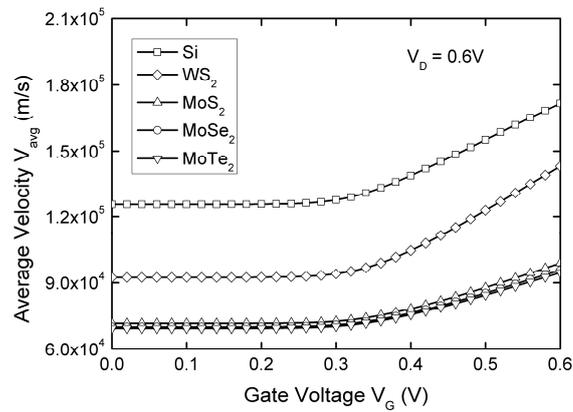

Fig. 6. Average velocity of carrier at the top of the energy barrier, $V_{avg}$, with increasing gate voltage, $V_G$, for n-type monolayer MX$_2$ transistors and 2D-Si transistors. Three sub-bands are considered in the calculation for Si thin film. $V_D$ is set at 0.6V.

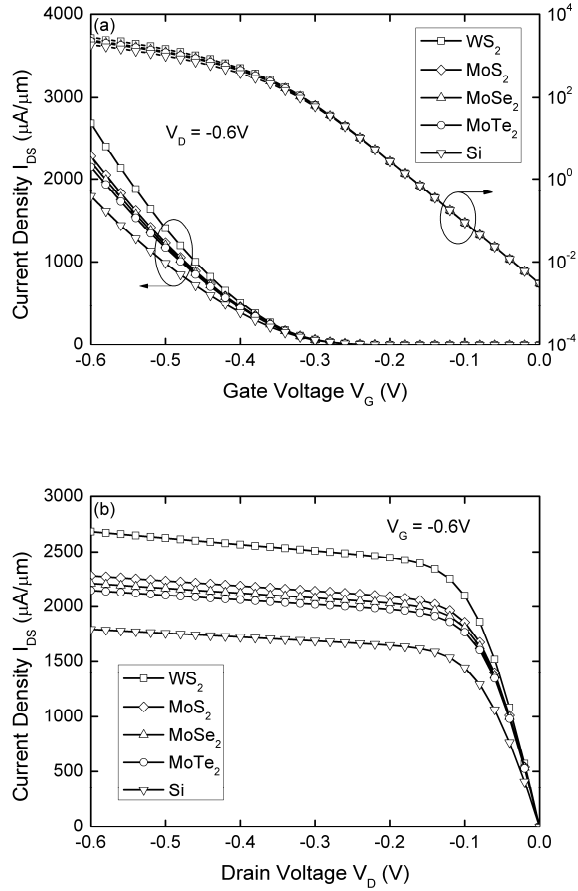

Fig. 7. Ballistic *I-V* characteristic comparison of p-type monolayer MX$_2$ transistors and 2D-Si transistors. Both the heavy hole and the light hole of Si are considered in the calculation. The off-current is set to be 0.003μA/μm. (a) $I_{DS} \sim V_G$ with $V_D$ = -0.6V. (b) $I_{DS} \sim V_D$ with $V_G$ = -0.6V.

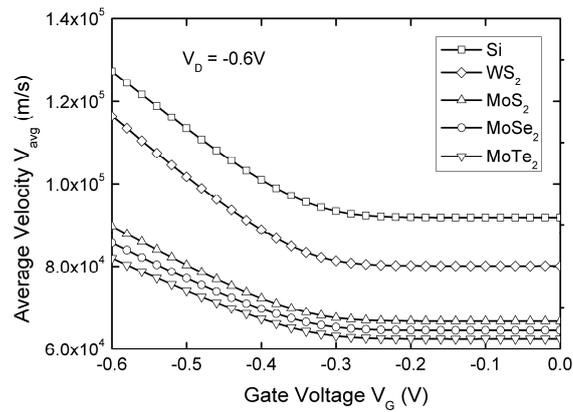

Fig. 8. Average velocity of carrier at the top of the energy barrier, $V_{avg}$, with increasing gate voltage, $V_G$, for p-type monolayer MX$_2$ transistors and 2D-Si transistors. Both the heavy hole and the light hole of Si are considered in the calculation. $V_D$ is set at -0.6V.